\documentstyle[aps,preprint]{revtex}
\begin{document}
\draft

\title{Viscosities of the Gay-Berne nematic liquid crystal}

\author{A. M. Smondyrev$^{1}$, George B. Loriot$^{2}$ and Robert A.
Pelcovits$^{1}$}

\address{
$^{1}$ Department of Physics, Brown University, Providence RI, 02912\\
$^{2}$ Computing and Information Services, Brown University, Providence RI,
02912}

\date{\today}
\maketitle

\begin{abstract}
We present molecular dynamics simulation measurements of the viscosities of the
Gay-Berne phenomenological model of liquid crystals in the nematic and
isotropic phases. The temperature-dependence of the rotational and shear
viscosities, including the nonmonotonic behavior of one shear viscosity, are in
good agreement with experimental data. The bulk viscosities are significantly
larger than the shear viscosities, again in agreement with experiment. \\
\end{abstract}
\pacs{61.30.-v, 61.30Cz,64.70Md}

\narrowtext

Ever increasing computer power has made simulations of simple, yet realistic
molecular models of liquid crystals a feasible enterprise. With the aid of
computers it is possible to study the effects of molecular shapes, sizes and
interactions on macroscopic behavior.Three types of liquid crystal models have
been used in simulation work. The first is the Lebwohl-Lasher model \cite{Leb},
 a lattice model for rotators. This model can be used to study the
isotropic-nematic transition  as a rotational order-disorder transition in an
effective crystalline solid. Another class of models which has received much
attention and includes the translational degrees of freedom of a liquid crystal
uses hard particles of various shapes which interact solely by excluded volume
effects \cite{Frenkel,Allen93}. These hard body models exhibit very rich phase
diagrams including smectic, columnar and cubatic phases. Finally, in recent
years there has been considerable numerical study of the Gay-Berne (GB) system
\cite{GB} which is a fluid  of point-like particles, each carrying a unit
vector ${\bf \hat u}$ which mimics the long molecular axis. The particles
interact via an anisotropic Lennard-Jones potential which depends on the
relative orientation and location of a pair of molecules. This system displays
rich behavior like the hard body models, but also includes attractive forces.
These latter forces play an especially important role in the formation of
smectic phases, which are more
readily formed in the GB system than in the hard body models..
The GB model has also been extended to include chirality\cite{Tsykalo} and to
model discotics\cite{Emerson}.

Previous studies of the GB system have focused on the phase diagram
\cite{deM,Luck90,LS93} and single-particle translational and rotational
dynamics\cite{DeRull}.  In this Letter we report on numerical measurements of
the viscosities of the GB system.  We find a number of results which indicate
that the GB model exhibits the principal dynamical features of a real nematic
liquid crystal. The temperature dependence of the shear viscosities  and the
rotational viscosity below the isotropic-nematic transition is qualitatively
similar to what is observed in experiment\cite{Mies,Langevin,Jeu,Ch}, including
the nonmonotonic behavior of one of the shear viscosities. The two bulk
viscosities are an order of magnitude larger than the shear viscosities in
agreement with ultrasonic measurements\cite{Letcher}. We also find that
long-time correlations of the director exhibit the expected Brownian motion due
primarily to the finite-size of our system.  This motion yields exceptionally
good statistics for the rotational viscosity.

The GB potential is modeled to give the best fit to the pair potential for a
molecula consisting of a linear array of four equidistant Lennard-Jones centers
with separation of $2\sigma_o$ between the first and fourth sites (subsequent
work\cite{LS93}  examined different fits) The GB potential is given by,
\begin{eqnarray}
\label{eq:U}
U({\bf \hat u}_1, {\bf \hat u}_2, {\bf r})&= & 4\varepsilon ({\bf \hat u}_1,
{\bf \hat u}_2, {\bf r})
\times \biggl[\biggl\{{\sigma_o\over r-\sigma({\bf \hat u}_1, {\bf \hat u}_2,
{\bf r}) +\sigma_o}\biggr\}^{12} \nonumber\\
& &-\biggl\{{\sigma_o\over r-\sigma({\bf \hat u}_1, {\bf \hat u}_2, {\bf
r})+\sigma_o}\biggr\}^6\biggr]
\end{eqnarray}
where ${\bf \hat u}_1, {\bf \hat u}_2$ are unit vectors giving the orientations
of the two molecules separated by the position vector $\bf r$. The parameters
$\varepsilon({\bf \hat u}_1, {\bf \hat u}_2, {\bf r})$ and $\sigma({\bf \hat
u}_1, {\bf \hat u}_2, {\bf r})$ are orientation dependent and give the well
depth and intermolecular separation where $U=0$ respectively.
The well depth is written as,
\begin{equation}
\label{eq:epsr}
\varepsilon({\bf \hat u}_1, {\bf \hat u}_2, {\bf r})=\varepsilon_o
\varepsilon^\nu ({\bf \hat u}_1, {\bf \hat u}_2)\varepsilon^{\prime\mu} ({\bf
\hat u}_1, {\bf \hat u}_2, {\bf r})
\end{equation}
where
\begin{equation}
\label{eq:epsnor}
\varepsilon({\bf \hat u}_1, {\bf \hat u}_2)=(1-\chi^2 ({\bf \hat u}_1 \cdot{\bf
\hat u}_2)^2)^{-1/2}
\end{equation}
and
\begin{eqnarray}
\label{eq:epsprime}
\varepsilon^\prime ({\bf \hat u}_1, {\bf \hat u}_2, {\bf \hat r})&= &
1-{\chi^\prime\over 2}\biggl\{ {({\bf \hat r}\cdot{\bf u}_1
+{\bf \hat r}\cdot{\bf u}_2)^2\over 1+\chi^\prime
({\bf u}_1\cdot {\bf u}_2)} \nonumber\\
& &+ {({\bf \hat r}\cdot {\bf u}_1-{\bf \hat r}\cdot{\bf u}_2)^2\over
1-\chi^\prime({\bf u}_1\cdot{\bf u}_2)}\biggr\}
\end{eqnarray}
The shape anisotropy parameter $\chi$ is given by
\begin{equation}
\label{eq:chi}
\chi=\{(\sigma_e/\sigma_s)^2 -1\}/\{({\sigma_e\over \sigma_s})^2+1\}
\end{equation}
where $\sigma_e$ and $\sigma_s$ are the separation of end-to-end and
side-by-side molecules respectively. The parameter $\chi^\prime$ is given by
\begin{equation}
\label{eq:chipr}
\chi^\prime
=\{1-(\varepsilon_e/\varepsilon_s)^{1/\mu}\}/\{1+(\varepsilon_e/\varepsilon_s)^{1/\mu}\}
\end{equation}
The ratio of the well depths for end-to-end and side-by-side  configurations
 is $\varepsilon_e/\varepsilon_s$.

We simulated a system of $N=256$ particles using the molecular dynamics
technique.  The ratio $ \sigma_{e} / \sigma_{s} $ was set equal
to 3, the ratio $ \epsilon_{e} / \epsilon_{s} = 5 $ and exponents
$\nu = 1$ and $\mu = 2$ . These values yield the best fit to the linear array
of four Lennard-Jones centers \cite{GB}, though other values have been used
\cite{Luck90,LS93}.  We used cubic periodic boundary conditions, and cut off
and smoothed the potential at  $ 3.8 \sigma_{0} $. The equations of motion were
solved
using the leap-frog algorithm with integration time-step
$ \Delta t^{*} = 0.001 $ in reduced units
 ( $\Delta t^{*} = \Delta t ( m \sigma_{0}^{2} / \epsilon_{0} )
^{-1/2}$, where $m$ is the mass of the molecule ). To  mimic the behavior of
$p$-azoxyanisole (PAA) we chose $\sigma_0=2.7\AA, \epsilon_0 / k_B = 400^\circ
K$, and $m=4.29 \times 10^{-23}$ g, yielding $\Delta t = 2.4 \times 10^{-12}$
sec. Translational and rotational temperatures were controlled using the
Nose-Hoover thermostat \cite{NH}. The initial
configuration was generated by locating nearly parallel GB molecules on the
sites of an fcc lattice at low scaled density, $ \rho^{*} (\equiv N \sigma_0^3/
V)= 0.10  $.
We carried out a long run (20,000 iterations) which reduced the nematic order
parameter to a value of 0.12, and disordered the system translationally. The
system was then gradually compressed at
constant scaled temperature $T^{*}(\equiv k_B T / \epsilon_0)=3.0$ to a density
$ \rho^{*} = 0.32 $.
Finally,  at constant density the temperature was lowered in small
steps to the isotropic-nematic transition temperature and below.  At this stage
 longer simulation runs
(180,000 - 300,000 iterations) were performed to obtain the values of the
viscosities.

The dynamical description of a compressible nematic requires six viscosities:
three shear viscosities, $\nu_1, \nu_2,$ and $\nu_3$; two bulk viscosities,
$\nu_4-\nu_2$, and $\nu_5;$ and a director rotational viscosity, $\gamma_1$. To
calculate these viscosities from correlation functions of the stress tensor and
the director, we transformed to a coordinate system where the 3-axis is
parallel to the average orientation of
the director, and the 1 and 2 axes
are perpendicular to the director.  The elements of the stress tensor are
defined by:
\begin{equation}
 \sigma_{\alpha \beta} = \frac{1}{V} ( \sum_{i} p^{i}_{\alpha}
p^{j}_{\beta} / m + \sum_{i} \sum_{j > i} r^{i j}_{\alpha}
f^{i j}_{\beta} ), \ \alpha,\beta=1,2,3,
\end{equation}
 where $V$ is the volume of the system, ${\bf p}^i$ is the linear momentum
of molecule $i$, and ${\bf r}^{ij}$ and ${\bf f}_{i j}$ are respectively the
relative position vector and force between molecules $i$ and $j$. The five
viscosities $\nu_1, \nu_2, \nu_3, \nu_4$, and $\nu_5$ associated with shear and
compression are then given in terms
of Kubo-like formulas \cite{Foster}:
\begin{eqnarray}
\label{eq:visco1}
\nu_{1}= \frac{V}{2k_{B}T} \int_{0}^{\infty} dt \; \lbrack<(\sigma_{33} (t) &-
&\sigma (t))( \sigma_{33}(0) - \sigma (0))> \nonumber\\
& &- <\sigma_{12} (t)  \sigma_{12}(0)> \rbrack
\end{eqnarray}
\begin{equation}
\nu_{2}  =   \frac{V}{k_{B}T} \int_{0}^{\infty} dt  \; <\sigma_{12} (t)
\sigma_{12}(0)>
\end{equation}
\begin{equation}
 \nu_{3}  =  \frac{V}{k_{B}T} \int_{0}^{\infty} dt  \; <\sigma_{13} (t)
\sigma_{13} (0)>
\end{equation}
\begin{equation}
 \nu_{4} = \frac{V}{k_{B}T} \int_{0}^{\infty} dt  \; <\sigma (t)  \sigma (0)>
\end{equation}
\begin{equation}
 \nu_{5} = \frac{V}{k_{B}T} \int_{0}^{\infty} dt \;  <\sigma_{33} (t) \sigma
(0)>
\end{equation}
where
\begin{equation}
 \sigma(t) = \frac{1}{2} ( \sigma_{11}(t) + \sigma_{22}(t))
\end{equation}

The time-correlation functions appearing in Eqs. (8)-(12) were evaluated by
averaging over successive time origins \cite{Allen-Til}. The components of the
stress tensor were evaluated first in the cubic coordinate system (with axes
parallel to the sides of the cell) where we defined periodic boundary
conditions  and then transformed to the director coordinate frame.

In order to compare our simulation results with experimental measurements of
shear viscosities, we applied a magnetic field
 $H$ along the long diagonal of simulation cube (the direction along which the
director spontaneously orders), with $\chi_a H^2 = 1.0$ in our reduced units,
($\chi_a$ is the anisotropy in the magnetic susceptibility). Experiments
typically measure Miesowicz viscosities \cite{Mies}:
\begin{equation}
\eta_1=\nu_3+{1\over4} \gamma_1 (1-\lambda)^2 + \lambda \gamma_1
\end{equation}
\begin{equation}
\eta_2=\nu_3+{1\over4} \gamma_1 (1-\lambda)^2
\end{equation}
\begin{equation}
\eta_3=\nu_2
\end{equation}
The parameter $\lambda$ is a reactive coefficient\cite{Foster} which determines
the response of the director to shear flow. Because we have not yet performed a
direct shear flow experiment we do not have a value of $\lambda$ for the GB
system; in PAA\cite{Foster} $\lambda=1.15\pm0.10$. We have assumed that
$\lambda$ is unity in the GB system and computed Miesowicz viscosities from our
measurements; a $10\%$ change in this value of $\lambda$ will not alter the
qualitative features of our results. Results for the three Miesowicz shear
viscosities are shown in Figs.~\ref{fig1} and ~\ref{fig2}; results for
$\gamma_1$ alone are given in Fig.~\ref{fig4}. The large value of $H$ needed to
stabilize the director motion smears the nematic-isotropic transition. At
sufficiently high temperatures where the nematic order parameter is low, (see
the inset of Fig.~\ref{fig1}), the system is essentially isotropic with large
director fluctuations and we have plotted the average of the two shear
viscosities $\eta_2$ and $\eta_3$ as a single isotropic viscosity. Below the
temperature $T\simeq 2.0$ where the nematic order parameter becomes substantial
we note that $\eta_2$
decreases at first as the temperature is lowered and then rises. This effect
has been observed experimentally in PAA\cite{Mies,Langevin},
N-($4^\prime$-methoxybenzilidene)-4-($n$ butyl) aniline (MBBA)\cite{Langevin},
and in the cyanobiphenyl homologues\cite{Ch}. The viscosity $\eta_2$ is
associated with shear flow parallel to the director, so its value drops when
the nematic order becomes appreciable. Its subsequent rise is not fully
understood. Likewise the fact that $\eta_3>\eta_2$ below
the transition has also been observed\cite{Mies,Langevin} and deep in the
nematic phase the ratio $\eta_3 / \eta_2$ has been found to be approximately
1.5, in both PAA and MBBA, in good agreement with our results. The temperature
dependence of $\nu_1$ or $\eta_1$ has not been measured, but the value of
$\nu_1$ deep in the nematic phase of PAA is known\cite{Letcher} and is
comparable to the value of $\eta_3$.
We find $\nu_1= 1.52$ and $1.044$ centipoise at $T^*=1.2$ and $1.4$
respectively. We have also measured the bulk viscosities at these latter
temperatures and find $\nu_4=19.6$ and $9.54$;$\nu_5=25.9$ and $10.74$
centipoise respectively.
 Experimental
measurements of the bulk viscosities deep in the nematic phase of PAA and
para-azoxyphenetole (PAP) \cite{Letcher} also show values which are an order of
magnitude larger than the largest shear viscosity. The temperature dependence
of the bulk viscosities has not been explored experimentally. Numerically
determining the bulk viscosities is more time consuming than determining the
shear viscosities, so we have not explored their full temperature dependence.

The magnitudes of our viscosities are of the correct order of magnitude for
what has been measured in PAA. The measured values of the viscosities for MBBA
are
consistently an order of magnitude higher than the corresponding quantities in
PAA. Simply changing the values of  $\epsilon_0, \sigma_0$, and the mass in the
GB potential to values appropriate to MBBA is not sufficient to account for
this difference. Nevertheless, the temperature dependence of the shear
viscosities is qualitatively similar in PAA and MBBA, so the present
parameterization of the GB potential captures the essential physics.

Correlations in the director rotational motion and the value of the rotational
viscosity $\gamma_1$ are best studied by measuring the correlation function:
\begin{equation}
\label{C}
C(t)=<(\delta n(t)-\delta n(0))^2>,
\end{equation}
where
\begin{equation}
\delta n(t) = \frac{1}{2} ( n_{1}(t) + n_{2}(t)),
\end{equation}
and,
\begin{equation}
n_{\alpha}(t) = \frac{1} {N}\sum_i u_{i \alpha}(t), \ \alpha = 1,2.
\end{equation}

If the director motion is diffusive (Brownian motion), then $C(t)$ at long
times will follow:
\begin{equation}
C(t)\sim 2{\gamma_1}^{-1}t,
\end{equation}
and the value of $\gamma_1$ can be extracted with excellent statistics. This
behavior was observed in our system (see Fig. ~\ref{fig3}) at sufficiently long
times, and is sensible given the magnetic field and especially the small system
size. At shorter times the director motion obeys the form $C(t) \sim t^2$
approximately, which is to be expected for ballistic behavior.  For much larger
system sizes and small magnetic fields we would expect this latter behavior to
persist to larger times due to the presence of massless Goldstone
modes\cite{Yeung}, even though the microscopic director motion is no longer
ballistic.  This Goldstone-like behavior has been reported by Zhang et
al.\cite{Zhang}, in Monte Carlo studies of the Lebwohl-Lasher lattice model
where very large system sizes are readily studied. The temperature
dependence of $\gamma_{1}$ is given in Fig.~\ref{fig4} in the
form of an Arrhenius plot, i.e., $\gamma_1 \propto \exp(E/k_B T)$, where $E$ is
an activation energy.  This Arrhenius behavior has been observed experimentally
 in MBBA and in a mixture of the two isomers of
$p$-methoxy-$p^\prime$-butylazoxybenzene (N4) \cite{Prost}. The slope of the
line give the activation
energy $E \simeq 0.1 eV$. This value has not been measured in PAA, but in MBBA
it is approximately 5 times larger. As discussed above, our numerical results
are consistently an order of magnitude lower in value than those measured in
MBBA. The value of $\gamma_{1}$ has been measured in PAA\cite{Gasp} deep in the
nematic phase ($T=122^\circ$) and the result,$\gamma_1 \simeq 6.8$ centipoise
is of the same order of magnitude as our results. \\

\acknowledgements
We are grateful to Prof. Seth Fraden for helpful suggestions. A.M.S. and R.A.P.
were supported in part by the National Science
Foundation under grant no. DMR92-17290.

\begin{figure}
\caption{The isotropic shear and two of the Miesowicz viscosities as functions
of the scaled temperature $T^*$, in the presence of a magnetic field. The inset
shows the nematic order parameter $S$ as a function of $T^*$ for the same value
of magnetic field. The parameters of the GB potential were chosen to mimic PAA;
see the text.}
\label{fig1}
\end{figure}

\begin{figure}
\caption{The third Miesowicz viscosity, $\eta_1$, which is two to three times
larger than the other two viscosities; compare Fig.\ \protect\ref{fig1}}
\label{fig2}
\end{figure}

\begin{figure}
\caption{The director correlation function $C(t)$, Eq.\ (\protect\ref{C}), as a
function of dimensionless time, at $T^*=1.4$. The short-time behavior is fit to
a straight line with slope $1.92$, while the slope at long times is $1.00$. The
error bars are imperceptible on this scale.}
\label{fig3}
\end{figure}

\begin{figure}
\caption{The rotational viscosity $\gamma_1$ as a function of scaled
temperature indicating the activated form of the dependence.}
\label{fig4}
\end{figure}

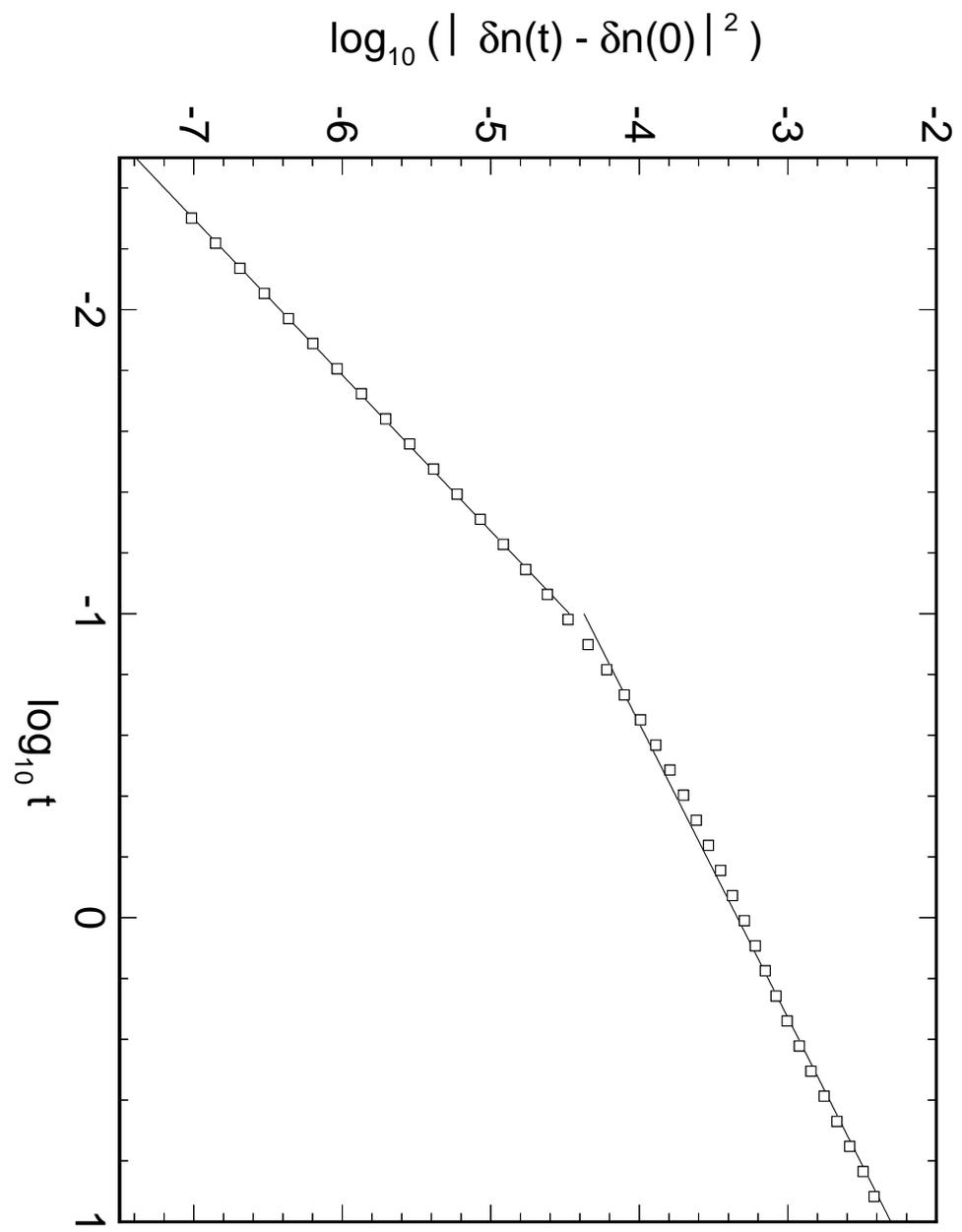
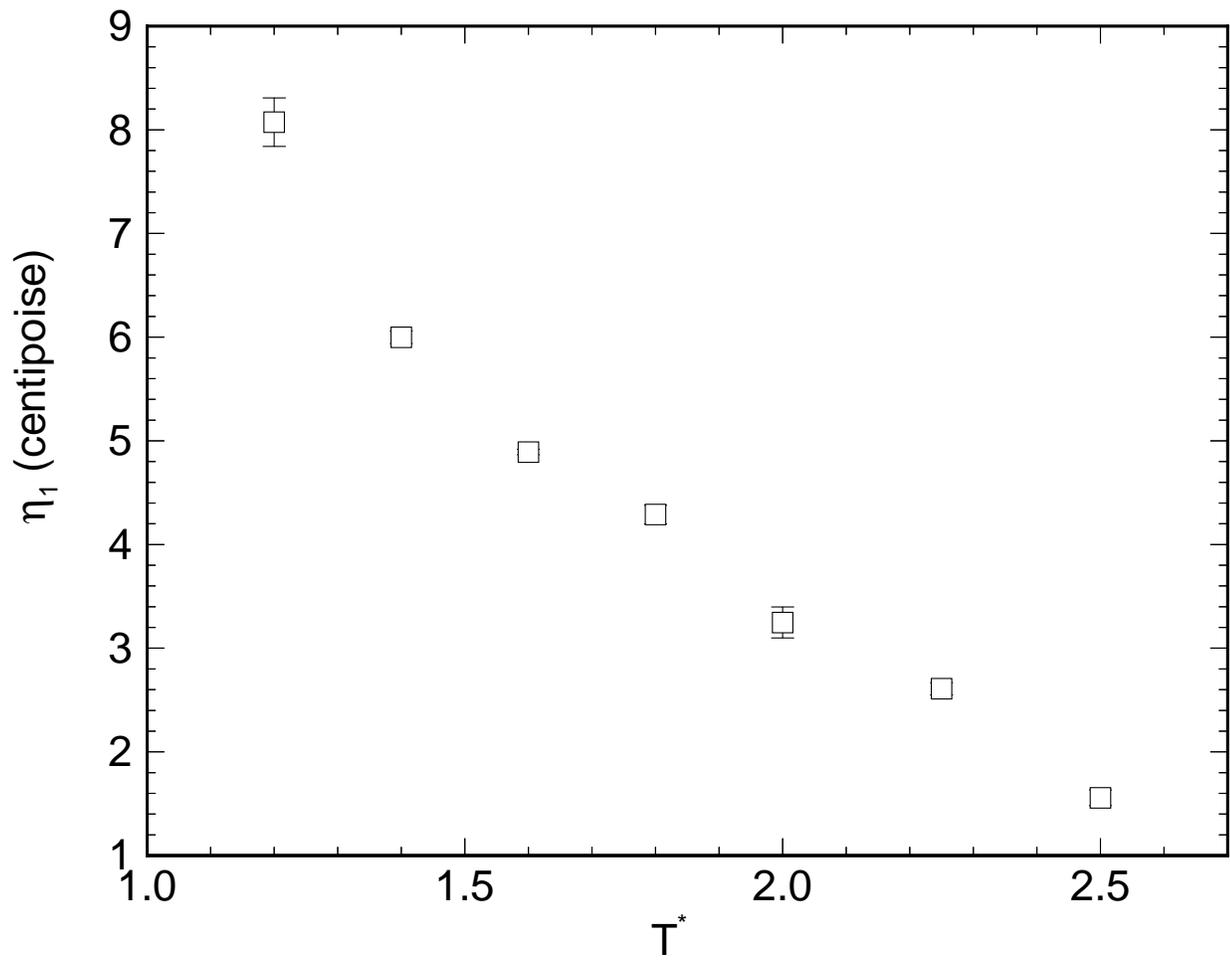
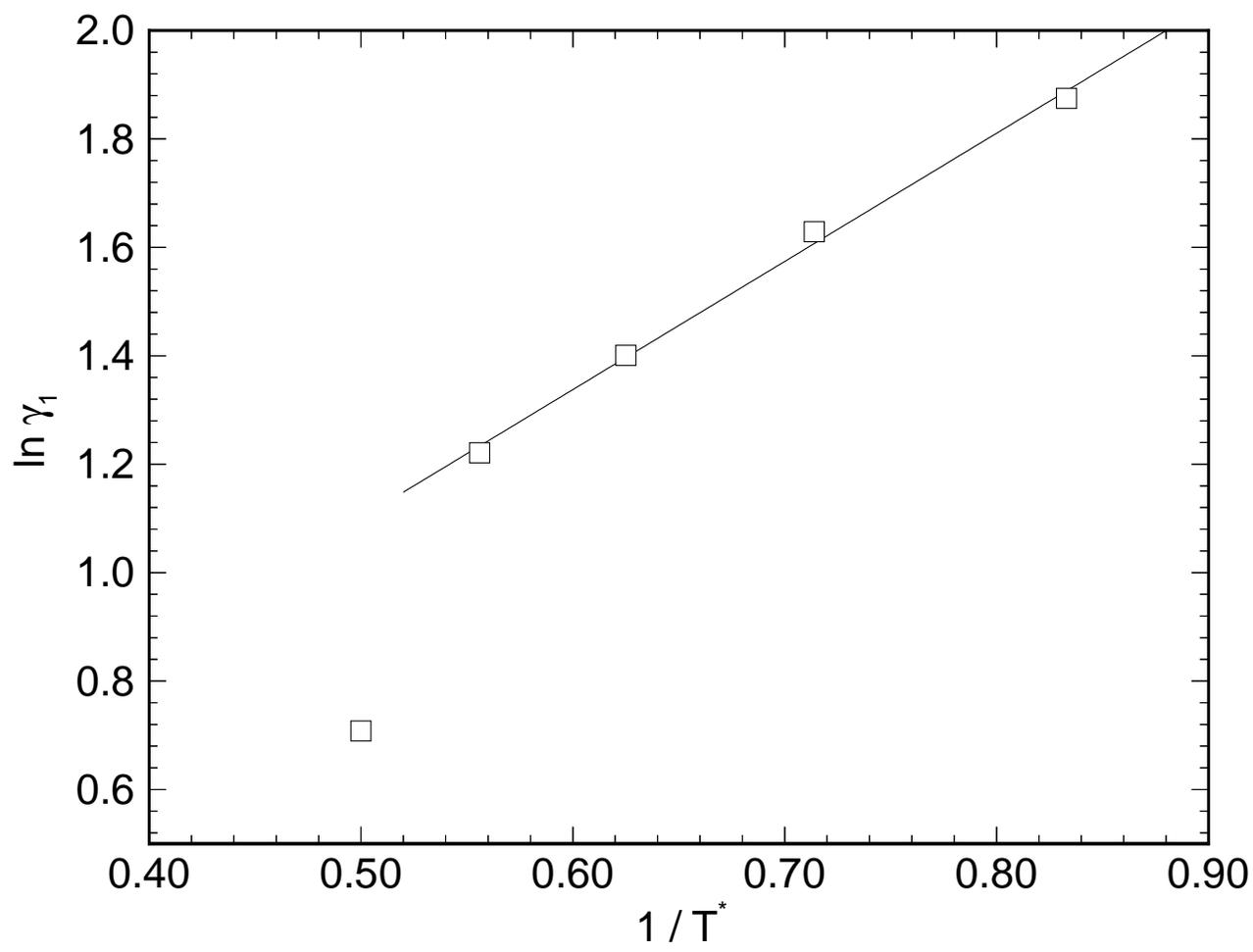
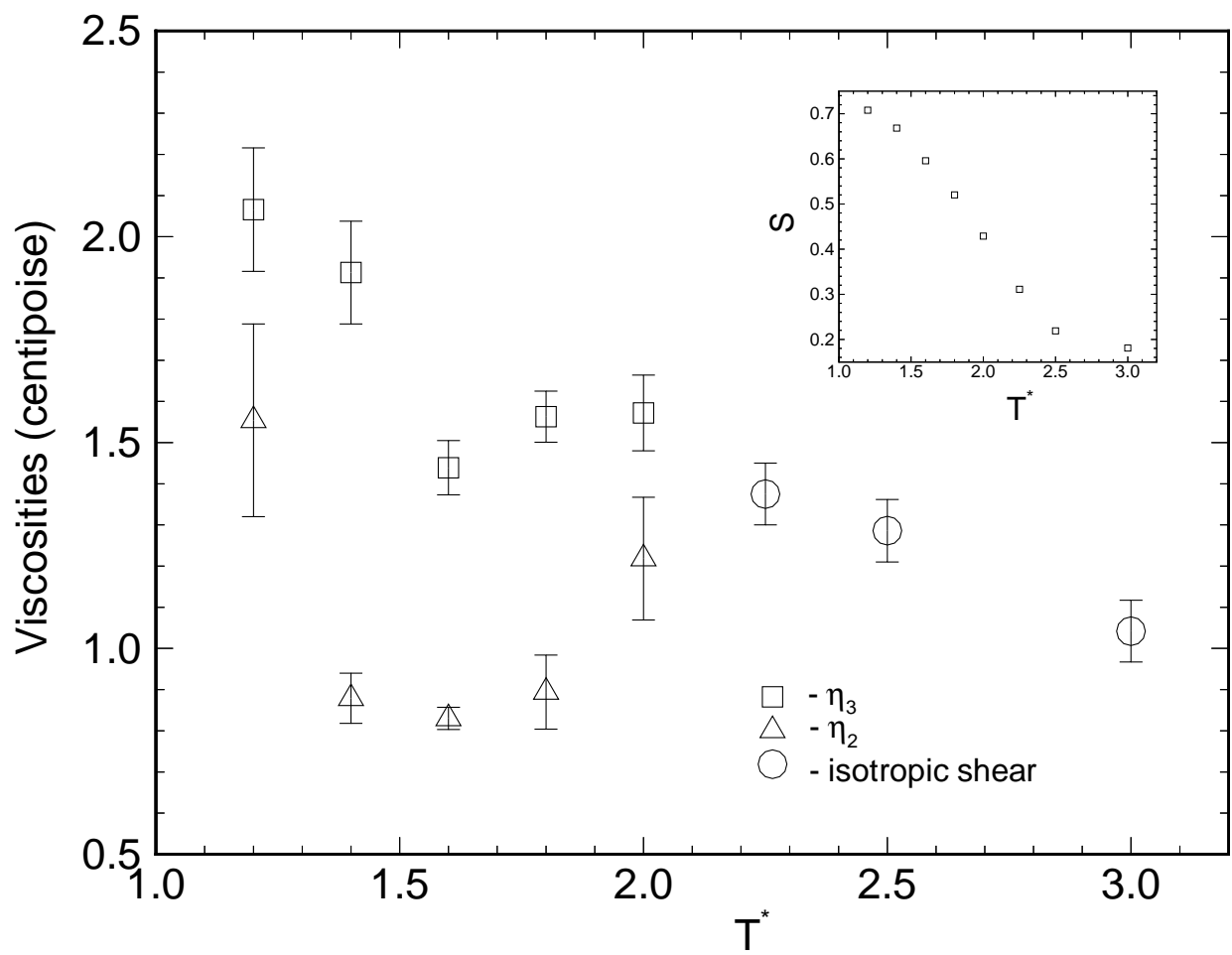
\end{document}